\documentclass[aps,prb,twocolumn,showpacs]{revtex4-1}
\usepackage{graphicx}
\usepackage{dcolumn}
\usepackage{color}
\usepackage{amsmath}
\newcommand{\qmc}{quantum Monte Carlo }

\begin{document}

\title{The Transition to the Metallic State in Low Density Hydrogen}

\author{Jeremy McMinis}
\author{Miguel Morales}
\affiliation{Lawrence Livermore National Laboratory, Livermore CA 94550}

\author{David Ceperley}
\affiliation{Department of Physics, University of Illinois, Urbana IL 61801}

\author{Jeongnim Kim}
\affiliation{Oak Ridge National Laboratory, Oak Ridge TN 37831}

\date{\today}

\begin{abstract}
Solid atomic hydrogen is one of the simplest systems to undergo a metal-insulator transition. Near the transition, the electronic degrees of freedom become strongly correlated and their description provides a difficult challenge for theoretical methods. As a result, the order and density of the phase transition are still subject to debate. In this work we use diffusion quantum Monte Carlo to benchmark the transition between paramagnetic and anti-ferromagnetic body centered cubic atomic hydrogen in its ground state. We locate the density of the transition by computing the equation of state for these two phases and identify the phase transition order by computing the band gap near the phase transition. These benchmark results show that the phase transition is continuous and occurs at a Wigner-Seitz radius of $r_s=2.27(3) a_0$. We compare our results to previously reported density functional theory, Hedin's GW approximation, and dynamical mean field theory results.
\end{abstract}

\maketitle

The metal-insulator transition in body centered cubic (BCC) hydrogen was first considered by Mott \cite{mott1958transition,mott1961transition}. At high density, the system is metallic: the electrons are in the paramagnetic, conducting phase. As the lattice spacing is increased the electrons localize on the hydrogen ions forming an anti-ferromagnetic, insulating phase. Predicting the order of the phase transition allows us to identify the qualitative nature of the transition: whether the transition is due to strong particle correlations, a Mott insulator transition, or due to long range Coulomb interactions, the band-insulator transition \cite{anisimov1991band,hubbard1963electron,imada1998metal,mott1990metal}. Locating the precise transition density is a difficult test for an electronic structure method. Since the original study, several closely related systems, finite lattices or 1D chains of hydrogen atoms, have become a standard test used to benchmark quantum 
chemistry methods \cite{Knowles1984315,doi:10.1021/ct301044e}. Unfortunately, current electronic structure methods such as density functional theory, dynamical mean field theory, variational quantum Monte Carlo, and Hedin's GW disagree over both the phase transition order and density \cite{Svane1990851,PhysRevLett.65.2414,PhysRevB.58.12680,PhysRevB.82.195123,PhysRevB.77.155114,PhysRevB.66.035102}.

In this work, we determine the phase transition density and order using diffusion quantum Monte Carlo (DMC). DMC is one of the most accurate methods for computing electronic structure properties of atomic, molecular, and condensed phases \cite{RevModPhys.73.33}. It has been used extensively to characterize hydrogen and the alkali metals at ambient and at high pressure \cite{PhysRevLett.105.086403,PhysRevLett.70.1952,holzmann2003backflow,Morales20072010}. We use DMC to compute the ground state equation of state in both the paramagnetic and anti-ferromagnetic phase as well as the quasiparticle and excitonic band gaps near the transition. 

We begin by reviewing previous calculations and the details of the current calculation. Next we discuss our results for the density of the phase transition and the magnitude of the band gap. Finally, we conclude with a comparison of results obtained with other methods and how they compare to those obtained using DMC.

\section{Previous Results}
The early calculations by Svane and Gunnarson showed that when self-interaction corrections were included in the local density approximation, density functional theory (DFT) predicted a first order phase transition located near the Wigner-Seitz radius $r_s=2.45 a_0$ where $r_s=(3/4\pi\rho)^{1/3}$, $\rho$ is the density, and $a_0$ is the bohr radius \cite{Svane1990851}. On the contrary, DFT calculations using either the generalized gradient approximation (GGA) or local spin density approximation (LSDA) without the self-interaction correction have predicted a second-order phase transition at $r_s=2.25 a_0$ and $r_s=2.5 a_0$ and an itenerant anti-ferromagnetic phase up to $r_s=2.5a_0$ and $r_s=2.8a_0$ respectively \cite{PhysRevB.58.12680}.

G$_0$W$_0$, using the LDA or GGA orbitals to compute the initial Green's function, finds the same transition order as their underlying DFT functionals, though the phase transition density is shifted upwards to $2.4<r_s<2.7$ \cite{PhysRevB.66.035102}. The most recent set of G$_0$W$_0$ calculations begin with LDA+U and GGA+U single particle orbitals for the initial Green's function \cite{PhysRevB.77.155114}. The ``+U'' methods include an on-site repulsion for the two different spin densities to penalize double occupancy and pushes the system towards an anti-ferromagnetic state. Using G$_0$W$_0$ on top of these methods, researchers find a continuous metal to insulator phase transition and locate it close to $r_s=2.2$.

This phase transition has also been investigated using dynamical mean field theory (DMFT) by approximating the Coulomb interaction as a strictly short ranged on-site interaction between two electrons on the same hydrogen ion \cite{RevModPhys.68.13,kotliar2004strongly,PhysRevB.82.195123}. Using this method it was found to be a first-order phase transition at $r_s\approx3$. This transition location is an extrapolation from their finite temperature data to the ground state \cite{PhysRevB.82.195123}.

A highly accurate benchmark is required to disambiguate these results. Previous efforts to produce such a benchmark have been performed using variational quantum Monte Carlo\cite{PhysRevLett.65.2414}. This calculation was consistent with either a very weak first order or a second order transition at $r_s=2.2a_0-2.3 a_0$. The error estimates in these measurements are sufficiently large to include a number of the previous results. Our goal in this work is to provide a benchmark with improved accuracy.

\section{Overview}
In this section we will discuss the method we use, the Hamiltonian for the system, and some computational aspects particular to our calculation.

\subsection{Diffusion Quantum Monte Carlo}
In this work we use DMC to generate all of our results. This method has been used to produce benchmark results for light elements such as hydrogen and the electron gas and has been increasingly used for solid state systems \cite{PhysRevLett.45.566, RevModPhys.84.1607}. This variational stochastic projector method filters out the ground state component of a trial wave function to sample the ground state probability distribution \cite{RevModPhys.73.33,reynolds:5593,PhysRevLett.71.2777}. By using a trial wave function we are able to avoid the notorious ``sign problem'' which plagues exact Monte Carlo calculations of fermions but introduce error which raises the energy. The nodes or phase of the trial wave function serves as a boundary condition on the the random walk. The error introduced by this approximation is referred to as the ``fixed-node error'' \cite{RevModPhys.73.33}.

\subsection{Hamiltonian}
In Rydberg units, the Hamiltonian for hydrogen is,
\begin{eqnarray}
 \hat{H} &=& -\sum_i\vec{\nabla}_i^2 + \sum_{i<j} \frac{2}{|r_{ij}|} - \sum_{I,j}\frac{2}{|r_{Ij}|} + \sum_{I<J}\frac{2}{|r_{IJ}|}
\end{eqnarray}
where capital letters, $I,J$, correspond to ion coordinates and lower case letter, $i,j$, correspond to electronic coordinates. This is a zero temperature calculation and does not include the kinetic energy of the  protons; they are clamped to the BCC lattice. In this work we will refer to the two atoms in the BCC unit cell as the A and B simple cubic sublattices.

\subsection{Trial Wave Functions}
Our trial wave function is a single Slater Jastrow wave function,
\begin{eqnarray}
 \Psi(\vec{R}) &=& \mathcal{J}_1(\vec{R})\mathcal{J}_2(\vec{R})\textrm{Det}^\uparrow(\vec{R})\textrm{Det}^\downarrow(\vec{R})
\end{eqnarray}
where $\vec{R}=\{ \vec{R}^\uparrow,\vec{R}^\downarrow \}$ where $\vec{R}^\uparrow = \{r^\uparrow_1,\ldots,r^\uparrow_{N^\uparrow}\}$ and similarly for the down spin electrons, $\vec{R}^\downarrow = \{r^\downarrow_1,\ldots,r^\downarrow_{N^\downarrow}\}$. For the ground state it is always the case that $N_\uparrow=N_\downarrow$. For the quasiparticle calculation they differ by 1. The Jastrow consists of two terms: a one-body term, $\mathcal{J}_1$, and a two-body term, $\mathcal{J}_2$\cite{PhysRev.98.1479} and are of  the form,
\begin{eqnarray}
 \mathcal{J}_1(\vec{R}) &=& \exp \left( \sum_{i,I} f^\sigma( |\vec{r}_i - \vec{r}_I | \right) \\
 \mathcal{J}_2(\vec{R}) &=& \exp \left( \sum_{i<j} g^{\sigma,\sigma^\prime}( |\vec{r}_i - \vec{r}_j | \right) 
\end{eqnarray}
where $I$ refer to ionic coordinates, $i,j$ refer to electron coordinates, $\sigma$ and $\sigma^\prime$ are the electron spins, and $f$ and $g$ are bspline\cite{spline} functions whose parameters are variational degrees of freedom. Both the one body and two body terms include a cusp condition which, in conjunction with the determinant, exactly cancels the divergent coulomb potential energy contribution when an ion and electron or two electrons coincide\cite{cusp}. We optimize the parameters in the trial wave function using a variant of the linear method of Umrigar and coworkers\cite{opt,toulouse:084102,Umrigar08}. Instead of rescaling the eigenvalues found during the generalized eigenvalue problem, we perform a line minimization on them using a $7$-point fit to a quadratic function. We find that this can increase the rate of convergence to the optimal set of variational parameters\cite{mmsd,mmsd2}.

We parameterize the two-body Jastrow function so that it is symmetric under exchange of up and down electron labels. This requires the same parameterization for $\mathcal{J}_2$ between up-up and down-down pairs, $g^{\uparrow,\uparrow}=g^{\downarrow,\downarrow}$, but allows for a separate set of parameters for up-down $\mathcal{J}_2$ terms,$g^{\uparrow,\downarrow}=g^{\downarrow,\uparrow}\neq g^{\uparrow,\uparrow}$.

The one-body Jastrow is parameterized differently in the paramagnetic and anti-ferromagnetic phases. In the paramagnetic phase we use a one body Jastrow which is not a function of electron spin or ion sublattice. In the anti-ferromagnetic phase we use a Jastrow that is the same for up-A/down-B, $f_A^\uparrow=f_B^\downarrow$, and for up-B/down-A, $f_B^\uparrow=f_A^\downarrow$, electron spin-ion sublattice pairs. This ensures that the wave function is unchanged if up and down electron labels are swapped at the same time as the A and B sublattice labels are.

For a Slater-Jastrow wave function, the magnitude of the fixed node error is affected by the quality of the single particle orbitals used in the Slater determinant. There are many different orbitals that can be used in the determinant: analytic forms such as plane waves or gaussians, orbitals from density function theory, or orbitals derived from another quantum chemistry method. It has been found that, for solids, the best trial wave functions are often obtained using orbitals from DFT. However, different functionals produce different orbitals, so a careful choice of functional is necessary \cite{PhysRevLett.101.185502, qmc-rmp}. 

In this work we consider single particle orbital sets generated using three different techniques all of which are based on DFT. In the paramagnetic phase we use orbitals generated using the local density approximation (LDA)\cite{LDA2}. In the anti-ferromagnetic phase, as we will describe below, we use two different sets of orbitals, one generated using a LDA+U functional, and the other generated by performing two LDA calculations, each of which includes only the ions of one sublattice. We generate these orbitals using Quantum Espresso \cite{QE-2009} and a LDA pseudopotential generated using OPIUM \cite{LDA2,opium:website}. 

\begin{figure}
 \includegraphics[width=0.96\linewidth]{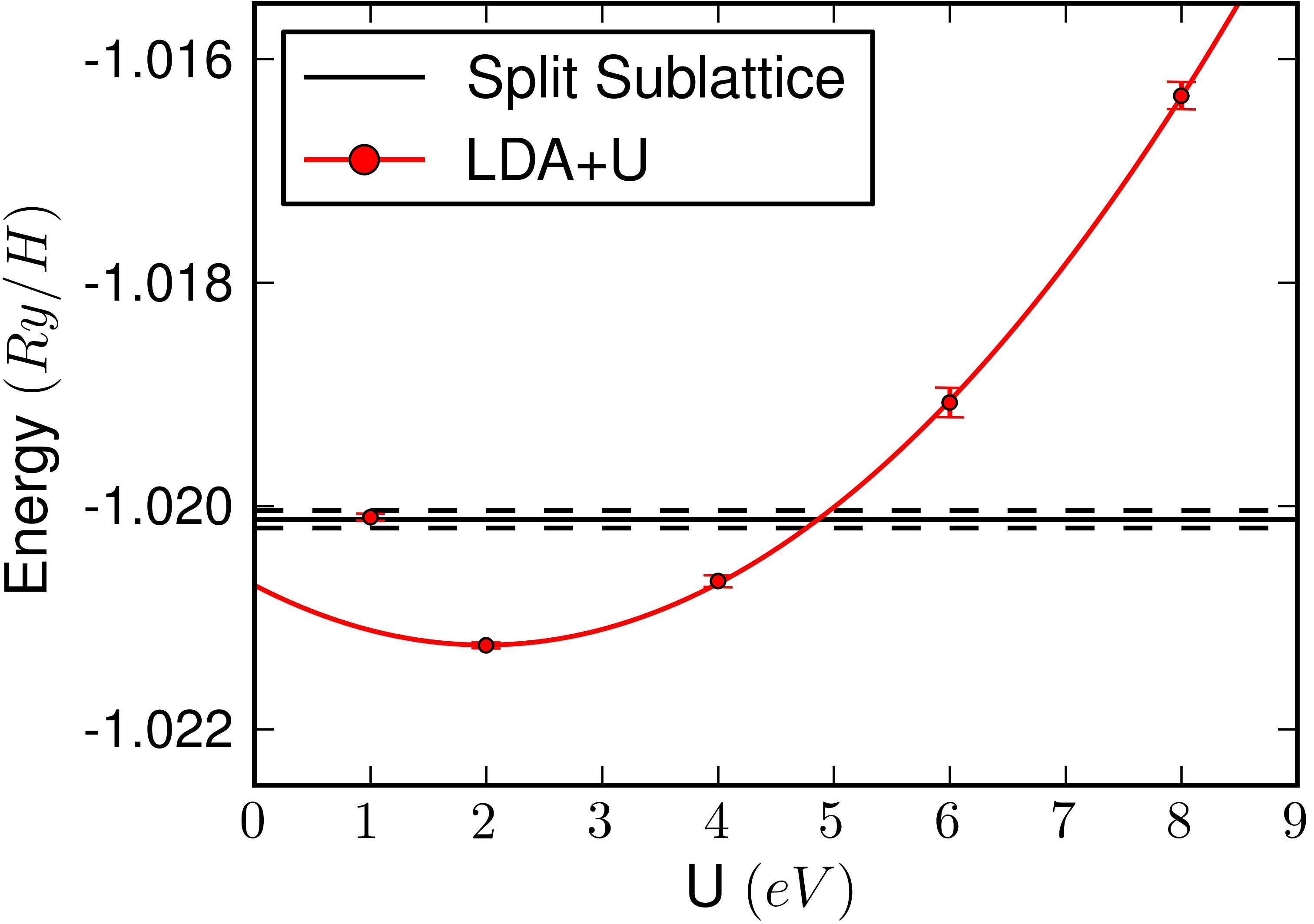}
 \caption{DMC Energy per hydrogen atom in Rydbergs for single particle orbitals generated using LDA+U. The X axis is the value of U used in the LDA+U DFT calculation ($1Ry = 13.6eV$). The horizontal line is the energy obtained using the split sublattice orbitals, and dashed line is a single standard deviation in energy. The minimum energy orbitals are generated using U$\approx2eV$. The line passing through the LDA+U energies is a quadratic fit for all U$\geq2 eV$.\label{FIG:POL_U}}
\end{figure}

\begin{figure}
 \includegraphics[width=0.96\linewidth]{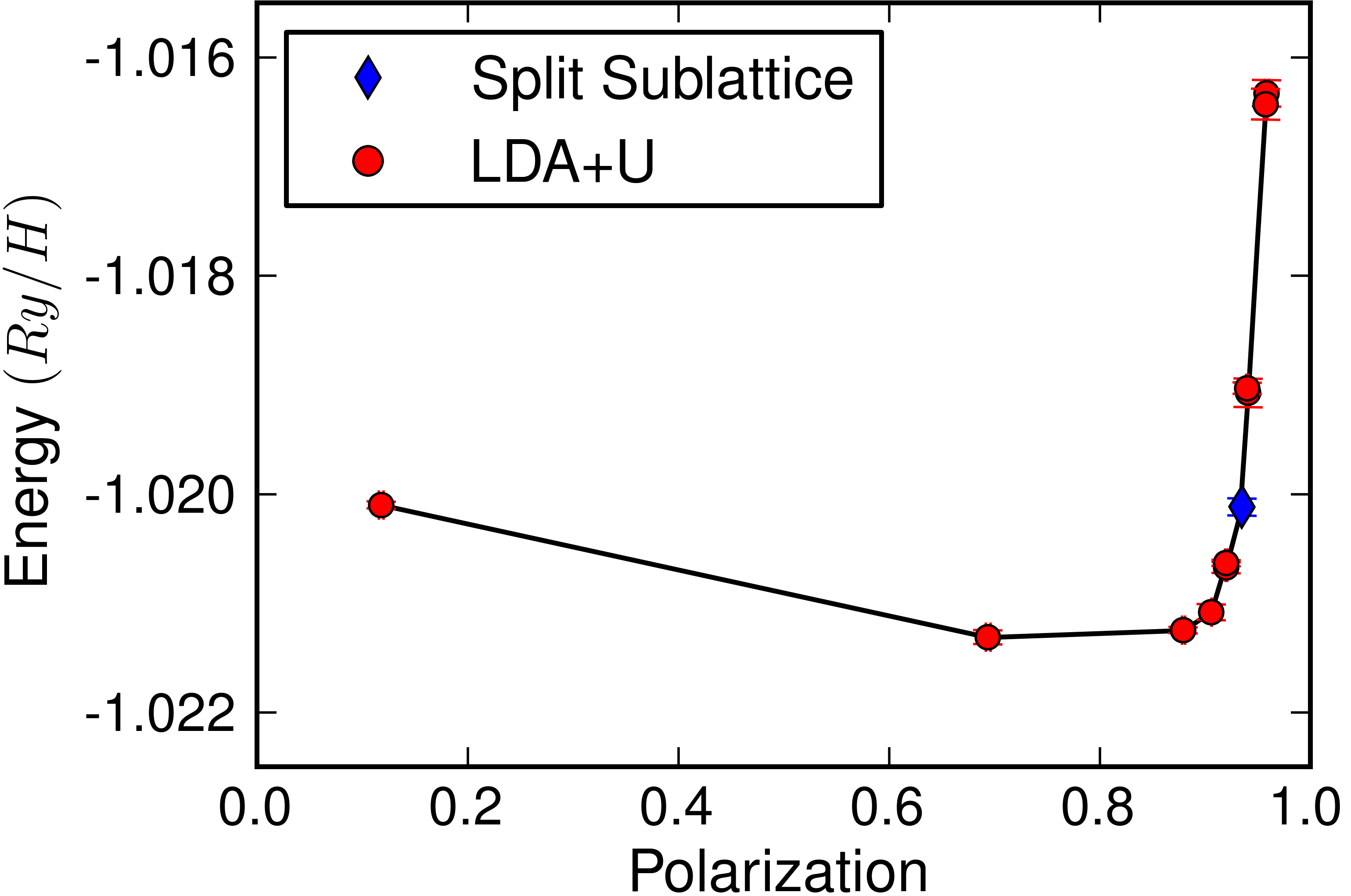}
 \caption{DMC Energy per particle in Rydbergs as a function of polarization at $r_s=2.3 a_0$. The minimum energy LDA+U orbitals (circles) are slightly less polarized than the split sublattice ones (diamond). The line is a guide to the eye.
\label{FIG:POL_E}}
\end{figure}

For the anti-ferromagnetic phase, we obtain the split sublattice orbitals by performing a DFT calculation on the simple cubic lattice and translating the orbitals according to the BCC basis vector,
\begin{eqnarray}
 \phi_A(\vec{R}) &=& \phi_B(\vec{R}+\vec{b}_1)\\
 \vec{b}_1 &=& \frac{l}{2}(1,1,1).
\end{eqnarray}
where $l$ is the lattice constant.
We choose to use this basis because, for the ground state, it explicitly breaks spin symmetry and creates a set of anti-ferromagnetic orbitals centered on each sublattice and has no adjustable parameters. 


We also use LDA+U orbitals to be able to tune the degree of spin polarization in the system. The LDA+U functional can interpolate between the paramagnetic and anti-ferromagnetic phases by tuning U. For small values of U the spin polarization is small and for large U it is complete. Though this is a simple single parameter single particle orbital optimization technique, it can be very expensive to converge. The sensitivity of the orbitals to the choice of U is illustrated in figure \ref{FIG:POL_U}. The resulting energy vs. polarization curve is presented in figure \ref{FIG:POL_E}. To find the optimal value of U we must perform QMC calculations on large systems, 432 electrons, for several values of U and orbital occupations. The orbitals may be occupied according to the LDA+U energy prediction, or by filling the lowest band regardless of LDA+U occupation. These partially polarized phases exhibit large finite size effects, and require large simulation cells to accurately determine the optimal U, its 
polarization and energy. Near the phase transition the LDA+U basis provides a better description of the anti-ferromagnetic state as evidenced by lower energies.


Because the DMC uses the full coulomb potential for the electron-ion potential, the only effect the choice of pseudopotential has on our DMC results is through the quality of the single particle orbitals. To check the effect of the pseudopotential, we compared the DMC energy of the single particle orbitals generated using our LDA pseudopotential with those generated using the Trail-Needs pseudopotential in the paramagnetic phase at $r_s=2.3 a_0$ \cite{trail:174109, trail:014112}. For the $2\times2\times2$ hydrogen supercell we found no statistical difference in their energies.

\subsection{Computing Polarization in Quantum Monte Carlo}
As described above, we use trial wave functions for the anti-ferromagnetic phase that explicitly break spin symmetry. We compute the polarization of the phase using the sublattice spin density,
\begin{eqnarray}
\langle \rho_A^{\sigma} \rangle &=& \left\langle \frac{1}{N_\sigma} \sum_{\substack{\vec{R}^A_I\in\{\vec{R}^A_1,\ldots,\vec{R}^A_N \}\\ \vec{r}^\sigma_i\in\{\vec{r}_1^\sigma,\ldots,\vec{r}_N^\sigma\}}} \Theta\left(|\vec r^\sigma_i - \vec R_I^A| - r_c\right) \right\rangle
\end{eqnarray}
where $\rho^\sigma$ is the spin density for the up or down electrons, $\sigma=\uparrow/\downarrow$, on the A sublattice whose protons are located at $\vec r_I^A$, and the step function is zero outside a sphere of radius $r_c$. $\rho_B$ is defined identically but with A and B indices exchange. We are free to adjust $r_c$ between zero and $r_{ws}$, the Wigner-Seitz radius. As $r_c\rightarrow r_{ws}$ the statistics of the observable improve due to increased sampling volume. As $r_c\rightarrow0$ the estimator projects out just the spin at the sublattice proton positions. We find that $r_c=0.5 a_0$ is a reasonable compromise and use it for all densities. Using these spin densities we define the A sublattice polarization, $S_A$, as,
\begin{eqnarray}
 S_A &=& \left|\frac{\langle\rho_A^{\uparrow} \rangle- \langle\rho_A^{\downarrow}\rangle}{\langle\rho_A^{\uparrow} \rangle + \langle\rho_A^{\downarrow}\rangle}\right|.
\end{eqnarray}
In this work we only investigate phases where the polarization is symmetric, $S_A=S_B=S$. In the paramagnetic phase $S=0$ and in the anti-ferromagnetic phase  $S\leq1$. This estimator works in trial wave function based QMC when the up/down symmetry of the state has explicitly been broken. If the trial wave function is a linear combination of up-A/down-B and up-B/down-A spin electrons assigned to sublattices, then the previous estimator yields zero even for anti-ferromagnetic phases. In that case it is necessary to compute the correlation function using the polarization operator,
\begin{eqnarray}
\hat{S} = \left\langle  (\rho_A^\uparrow - \rho_A^\downarrow) (\rho_B^\downarrow - \rho_B^\uparrow) \right\rangle.
\end{eqnarray}

\subsection{Computational Details}
We converge the DFT calculations using a $24\times24\times24$ K-point grid for the metallic phases and $12\times12\times12$ grid for the insulating or partially polarized phases.  We find that using an energy cutoff of $360/r_s$ Rydberg is sufficient to converge the DFT energy as well as the orbitals used in the quantum Monte Carlo. 

In the paramagnetic phase, we perform DMC for a $3\times3\times3$ supercell of $8\times8\times8$ twists and $54$ total ions, and a $4\times4\times4$ supercell of $6\times6\times6$ twists and $128$ total ions using both the Ewald and modified periodic Coulomb (MPC) potential \cite{PhysRevB.78.125106} for the electrons \cite{PhysRevB.53.1814,PhysRevE.64.016702}. These runs were performed with more than two thousand configurations and extrapolated to zero time step using a three point extrapolation. The smallest time steps had accept ratios greater than $99.5\%$. The energy is then averaged over twist points and the two supercells are extrapolated linearly as Volume$^{-1}$ to infinite volume. We find that both the Ewald and MPC extrapolate to similar values, and forgo larger simulation cells in the extrapolation.  

We put more effort into computing the properties of the anti-ferromagnetic phase because it is relatively more complex than the paramagnetic phase. Because the anti-ferromagnetic phase is insulating,  we use a smaller number of twists. The finite size and Brillouin zone sampling are done using a $3\times3\times3$ supercell of $4\times4\times4$ twists ($54$ protons), a $4\times4\times4$ supercell of $3\times3\times3$ twists ($128$ protons), and a $6\times6\times6$ supercell of $2\times2\times2$ twists ($432$ protons).

\section{Phase Transition Density}
We locate the transition density between the paramagnetic and anti-ferromagnetic phases by computing the energy of each phase as a function of Wigner-Seitz radius and finding the crossing point. By carefully extrapolating finite size effects, sampling the Fermi surface using twist averaging, and controlling for time step error, we determine the transition density within a small statistical error \cite{PhysRevE.64.016702}. We control for systematic error, due to the fixed node approximation, by comparing the energies for several different sets of single particle orbitals in the anti-ferromagnetic phase and by using a Slater-Jastrow-backflow wave function in the paramagnetic phase.

\subsection{Results}
\begin{table}
\centering
\begin{tabular}{|c c c |}
\hline
$r_s (a_0)$ & AFM & PM \\\hline
1   &-0.67276(5) & \\
2   &-1.02309(5) & \\
2.1 &-1.02211(11) &-1.03873(2)\\          
2.3 &-1.01955(5) &-1.01948(2)\\
2.4 &-1.01705(8) &\\
2.5 &-1.01503(3) &-1.00078(2)\\           
2.7 &-1.01072(2) &-0.98501(3)\\           
3   &-1.00616(2) & \\                                 
4   &-1.00064(3) & \\
5   &-0.99971(1) & \\\hline
\end{tabular}
\caption{Fixed-phase DMC Energy per hydrogen atom in Rydberg for BCC hydrogen in the anti-ferromagnetic (AFM), and paramagnetic (PM) phases.\label{TBL:H_EOS}}
\end{table}

\begin{figure}
\centering
\includegraphics[width=0.96\linewidth]{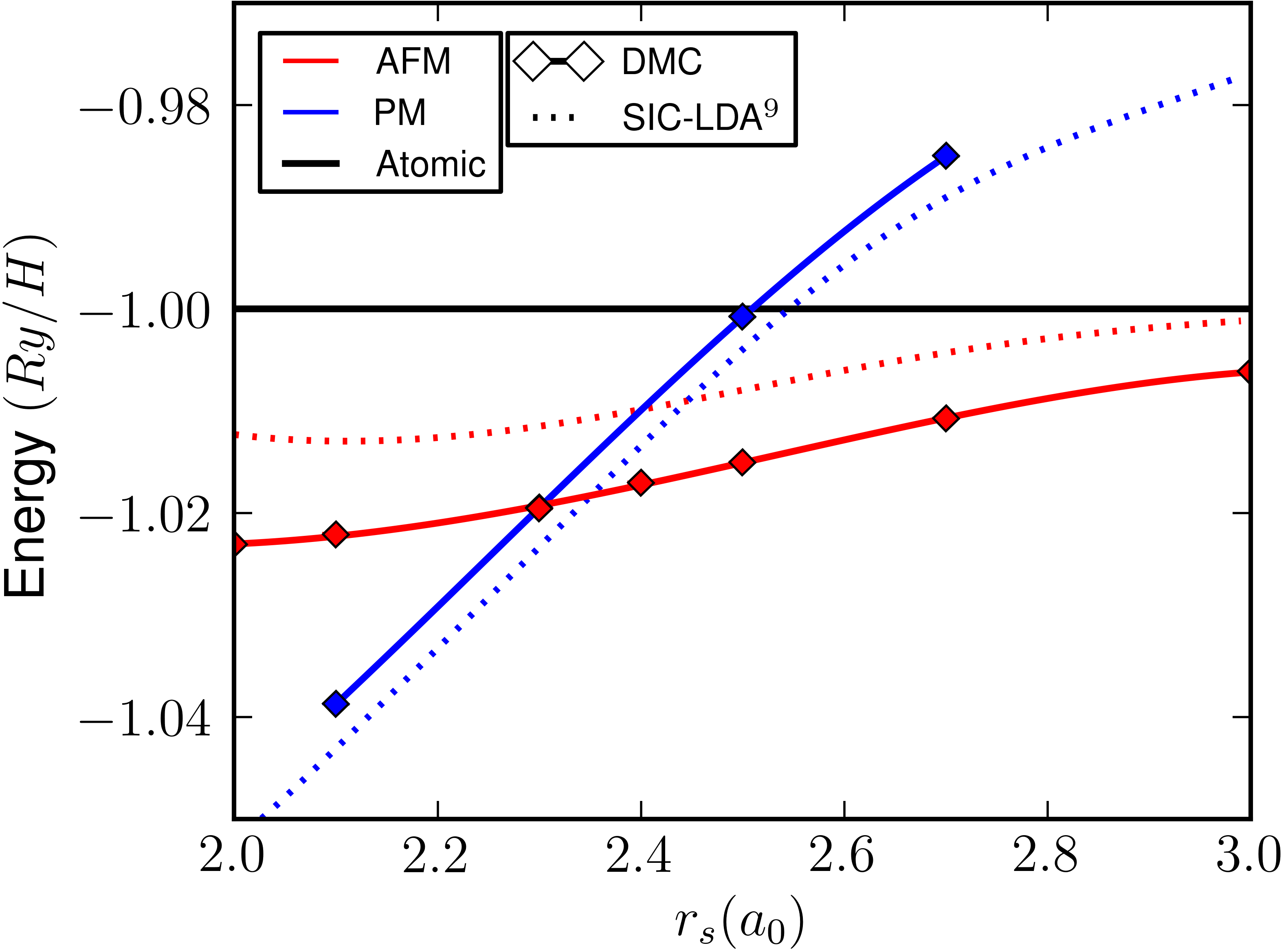}
\caption{Energy per atom in Rydberg for BCC hydrogen in the anti-ferromagnetic, and paramagnetic phases. The diamonds and solid line are DMC results. The error bars are smaller than the symbols. The anti-ferromagnetic and paramagnetic energies at $r_s=2.3$ are coincident. The energy of the anti-ferromagnetic phase asymptotes to the atomic limit as $r_s\rightarrow\infty$.
\label{FIG:EOS_trans}
}
\end{figure}

We present results for the equation of state of the anti-ferromagnetic phase at $r_s=1-5$ and the paramagnetic state near the phase transition in table \ref{TBL:H_EOS}. These results are from fully converged DMC calculations. The data presented in table \ref{TBL:H_EOS} are plotted in figure \ref{FIG:EOS_trans} near the phase transition. We also provide the self-interaction corrected DFT (SIC-DFT) results for reference. 

These equations of state were calculated at the DMC level using the split sublattice orbitals in the anti-ferromagnetic phase and LDA orbitals in the paramagnetic one. When reporting the transition density we must also include an estimate of the systematic error. Improvements in the quality of the wave function in the paramagnetic phase will push the phase transition to slightly higher $r_s$ while improvements to the anti-ferromagnetic phase will push it lower. 

We use more accurate trial functions to estimate our systematic error at $r_s=2.3a_0$. Near the transition on the anti-ferromagnetic side, we present the energy of trial wave functions using LDA+U single particle orbitals as a function of U in figure \ref{FIG:POL_U}. The energy gain using the best LDA+U single particle orbitals in the Slater-Jastrow wave function is around $1.5 mRy/N$. The statistical uncertainty in the shift is much smaller $\approx 0.01 mRy/N$. In the paramagnetic phase, using a Slater-Jastrow-backflow trial wave function, near the transition density we find an energy difference of less than $0.5 mRy/N$. We use these energy shifts to move the location of the phase transition and also as an estimate of the total systematic error. When these systematic errors are included it lowers the transition Wigner-Seitz radius $0.02a_0$. This puts the phase transition at $r_s=2.27(3)$.






%

\begin{table}
\centering
\begin{tabular}{|c c c|}
\hline
Method & transition $r_s$ & order\\ \hline
DMC  & $2.27(3)$ & 2 \\\hline
G$_0$W$_0$/LDA+U \cite{PhysRevB.77.155114}& 2.24 & 2\\
G$_0$W$_0$/GGA+U \cite{PhysRevB.77.155114}& 2.21 & 2\\
G$_0$W$_0$/LSDA \cite{PhysRevB.77.155114} & 2.65 & 2\\
G$_0$W$_0$/GGA \cite{PhysRevB.77.155114} & 2.42 & 2\\
VQMC \cite{PhysRevLett.65.2414} & 2.25(5) & 2?\\
LSDA \cite{PhysRevB.58.12680} & 2.78 & 2\\
GGA \cite{PhysRevB.58.12680} & 2.50 & 2\\
SIC-LSDA \cite{Svane1990851} & 2.45 & 1\\
DMFT \cite{PhysRevB.82.195123} & $\sim$3.0 & 1\\ \hline
\end{tabular}\\
 \caption{ Comparison for the transition Wigner-Seitz radius, $r_s$, and phase transition order with previous methods. The DMFT results are using the Hubbard model instead of the Coulomb potential and are extrapolated from finite temperature to zero for comparison here. \label{TBL:HEOS_prev}}
\end{table}

\section{Phase Transition Order}
Because quantum Monte Carlo methods do not provide direct access to the density of states or excitation spectra, we identify the phase transition order by investigating the band gap of the system near the phase transition density. A finite gap will not allow us to rule out a weak first order transition, but a gapless system indicates that the phase transition is continuous. We compute two different band gaps, the quasiparticle and excitonic band gap.


The excitonic band gap is the energy necessary to create a particle-hole excitation\cite{PhysRevLett.72.2438}. We compute this gap by computing the energy of the lowest lying exciton and subtracting off the ground state energy,
\begin{eqnarray}
 \Delta E_{\textrm{exc}} &=& E_1-E_0.
\end{eqnarray}
where
\begin{eqnarray}
 |\Psi_{1}\rangle&=& a_{LU}^\dagger a_{HO} |\Psi_{0}\rangle\\
 E_1 &=& \langle \Psi_{1} | \hat H |\Psi_{1}\rangle
\end{eqnarray}
and $a_{LU}^\dagger$ creates a particle in the lowest energy unoccupied orbital and $a_{HO}$ destroys a particle in the highest energy occupied orbital\cite{kent1998quantum,PhysRevLett.72.2438}. In this case, the lowest lying exciton is an indirect exciton. The particle state is at the $X$ point and the hole state is at the $R$ point in reciprocal space.
This does not result in any additional complications for our \qmc calculations other than requiring that both reciprocal space points be commensurate with the supercell.

The quasiparticle gap is different from the excitonic gap in that it does not include the exciton binding energy. It is computed as the difference in energy between a quasi-electron and an quasi-hole state,
\begin{eqnarray}
 \Delta E_{\textrm{qp}} &=& (E_{N+1}-E_{N}) - (E_N-E_{N-1})
\end{eqnarray}
where $E_N$ is the ground state energy of the system with $N$ electrons. This requires an additional neutralizing background charge, which can cause and additional finite size error \cite{kent1999techniques,kent1998quantum}. If the total number of electrons is increased or decreased by one, a background charge must be included or the overall electrostatic energy of the cell will diverge. In practice we accomplish this by ignoring the $k=0$ component of the electron-ion and electron-electron Coulomb interaction. Because this is a charged cell, we use the Makov and Payne extrapolation, $1/L$, to the thermodynamic limit \cite{PhysRevB.51.4014}. 

We compute the band gaps using both methods using our best LDA+U single particle orbitals at $r_s=2.3$. This location is very close to the phase transition. For these gaps we use $6\times6\times6$ and $4\times4\times4$ supercells and extrapolate to the infinite limit. More than two thousand configurations were used and the time step extrapolation is handled the same as for the equation of state calculations. We find a very small band gap which is consistent with a transition density closer to $r_s=2.3 a_0$ than methods considered in Table \ref{TBL:HEOS_prev}.

\begin{table}
\centering
\begin{tabular}{|c c c|}
\hline
Method & supercell & Gap (eV)\\ \hline
Excitonic & $4\times4\times4$ & 0.38(7)\\
 & $6\times6\times6$ & 0.1(1)\\
as $1/N$ & $\infty\times\infty\times\infty$ & 0.0(1) \\ 
Quasiparticle & $4\times4\times4$ & -0.54(5)\\ 
 & $6\times6\times6$ & -0.24(5)\\
as $1/L$ & $\infty\times\infty\times\infty$ & 0.36(18)\\ \hline
G$_0$W$_0$:GGA+U\cite{PhysRevB.77.155114} & & $\approx 1.2$\\
G$_0$W$_0$:LDA+U\cite{PhysRevB.77.155114} & & $\approx 2.0$\\
\hline
\end{tabular}\\
 \caption{ Quasiparticle and Excitonic band gaps at $r_s=2.3 a_0$. \label{TABLE:Gaps}}
\end{table}

The gaps we compute and reference values from previous studies are presented in table \ref{TABLE:Gaps}. The excitonic gap extrapolates to zero gap while the quasiparticle gap projects to a small but finite gap, $0.36(18) eV$. Both of these gaps are consistent with a phase transition location closer to $r_s=2.3 a_0$ than previous G$_0$W$_0$ studies. We note that the quasiparticle gap suffers from much larger finite size effects than the excitonic gap which results in a more difficult extrapolation and larger error bar estimate. At this density, an estimate for the excitonic binding energy, the difference between the excitonic and quasiparticle gap extrapolated to the thermodynamic limit, is $0.36(10) eV$.

\section{Conclusions}
We have presented benchmark calculations on the phase transition density and order for BCC hydrogen. As shown in table \ref{TBL:HEOS_prev}, the phase transition is likely to be continuous and is located at $r_s=2.27(3)a_0$. The most recent G$_0$W$_0$ calculations\cite{PhysRevB.77.155114} show the same transition order and agree very well with our transition location. If these calculations were to choose a slightly different value of U, they may be brought into perfect agreement. Previous publications\cite{PhysRevB.66.035102} using G$_0$W$_0$ resulted in slightly less accurate transition densities, likely due to a worse DFT starting point. Our results provide confidence in the G$_0$W$_0$ method when starting from a high quality DFT calculation. Previous DFT results identify the same phase transition order, but greatly overestimate the transition location\cite{PhysRevB.58.12680}. 
Previous QMC results\cite{PhysRevLett.65.2414} were performed on small systems, of $54$ ions, which is under converged with respect to system size. They are also performed using VMC which is much more sensitive to the form and optimization of the trial wave function. However, these results predict a similar transition density as the current ones. The remainder of the methods, DMFT \cite{PhysRevB.82.195123} and SIC-DFT \cite{Svane1990851} get the phase transition order and location incorrect.



\section{Acknowledgements}
JM acknowledges useful conversations with Sarang Gopalakrishnan, Norm Tubman, and Lucas Wagner. This work was performed in part under the auspices of the U.S. Department of Energy (DOE) by LLNL under Contract DE- AC52-07NA27344. JM and DMC were supported by NSF OCI-0904572. JM, DMC, MM, and JK were supported by the Network for ab initio many-body methods supported by predictive theory and modeling program of Basic Energy Sciences, Department of Energy (DOE). This work used the Extreme Science and Engineering Discovery Environment (XSEDE), which is supported by National Science Foundation grant number OCI-1053575, and resources provided by the Innovative and Novel Computational Impact on Theory and Experiment (INCITE) at the Oak Ridge Leadership Computing Facility at the Oak Ridge National Laboratory, which is supported by the Office of Science of the U.S. Department of Energy under Contract No. DE-AC05-00OR22725. LLNL release number LLNL-JRNL-638420.


%

\end{document}